\documentclass{article}

\usepackage{arxiv}
\usepackage{amsthm}
\usepackage[utf8]{inputenc} 
\usepackage[T1]{fontenc}    
\usepackage{hyperref}       
\usepackage{url}            
\usepackage{booktabs}       
\usepackage{amsfonts}       
\usepackage{nicefrac}       
\usepackage{microtype}      
\usepackage{lipsum}		
\usepackage{graphicx} 
\usepackage{subfigure}
\usepackage{float}

\title{A Coding-free Software Framework of Developing Web Data Management Systems}

\author{
  Can Yang \\
  Computer Science and Engineering\\
  South China University of Technology\\
  Guangzhou,China \\
  \texttt{cscyang@scut.edu.cn} \\
   \And
 Shiying Pan \\
  Computer Science and Engineering\\
  South China University of Technology\\
  Guangzhou,China \\
  \texttt{201721041376@mail.scut.edu.cn} \\
   \And
  Runmin Li \\
  Computer Science and Engineering\\
  South China University of Technology\\
  Guangzhou,China \\
 \And
  Yu Liu \\
  Shenzhen Tencent computer systems co.,LTD.\\
  Shenzhen, China\\
 \And
  Lizhang Peng\\
  Vipshop Holdings Limited\\
 Guangzhou, China\\
}

\begin{document}
\maketitle

\begin{abstract}
More and more enterprises recently intend to deploy data management systems in the cloud. Due to the professionalism of software development, it has still been difficult for non-programmers to develop this kind of systems, even a small one. However, the development of SaaS brings forth the more feasibility of coding-free software development than before. Based on the SaaS architecture, this paper presents a set of theory and method for coding-free construction of a data management system, on which our contributions involve in a practical application platform, a set of construction method and a set of interface on data exchange. By abstracting the common features of data management systems, we design a universal web platform to quickly generate and publish customized system instances. Moreover, we propose a kind of method to develop a data management system using a specific requirements table in spreadsheet. The corresponding platform maps the requirements table into a system instance through parsing the table model and implementing the objective system in the running stage. Finally, we implement the proposed framework and deploy it on web. The empirical result demonstrates the feasibility and availability of the coding-free method in developing web data management systems.
\end{abstract}

\section{Introduction}
In recent decades, the concept of software engineering has become increasingly mature. It's well-known that coding is of necessity in the software development process. Due to the strong professionalism of software development, it often takes a lot of time for one without programming experience to learn coding and develop a software. We know that data management is the core of most information systems, in which there are actually a number of common requirements in the development of the systems. However, it is often impossible for non-programmers\cite{Rosson05} to develop a data management software by themselves. In this paper, we provide possibilities for non-programmers to develop software systems and show them how to do so.

The classic pattern of software development is that programmers who are familiar with coding serve a wider audience\cite{Christopher05}. However, the number of users without programming knowledge is much more than that of programmers. Spreadsheets are ubiquitous information management tools cross technical and cultural divides. According to the 2012 United States Census, over 55 million people in the United States ``manipulate data with either spreadsheets or databases at work'' \cite{Benson14}. Therefore, we try to simplify the software development by using the ease of use of spreadsheets.

Based on the above-mentioned reasons, this paper proposes the ideas of coding-free software development. Our purpose is to design a web application software architecture by extracting common features in similar software requirements, and to implement an agile software development framework supported requirements-table-driven.

Today, cloud-oriented SaaS provides a viable technical approach for coding-free software development. SaaS model refers to that enterprises or individuals use software hosted and managed by SaaS providers\cite{Mietzner08}, who aim to bring economic benefits by taking advantage of the commonality of large-scale software\cite{Guo07}. Under the SaaS model, enterprises no longer have to buy, build and maintain infrastructure. According to IDC(International Data Corporation), the global cloud software market will grow at a compound annual rate of 18.3 percent(CAGR) over \$112.8 billion by 2019. SaaS delivery will significantly exceed traditional software product delivery, and the growing nearly five times faster than the traditional software market.

In the reminder of the paper, we introduce a general development framework of data management systems based on SaaS. The framework involves such three aspects as a Data Management System Web Platform (DMSWP), a system Requirements-Table-Driven Development Method (RTDDM), and a set of General Data Web Interfaces (GDWI). Also, our contributions lies in them as follows:
\begin{itemize}
\item DMSWP: The platform provides a scalable, multi-tenant software application model, which provides both tenants(system administrators) and ordinary users accessing to the system. The platform supports tenants to develop software systems based on RTDDM, and also supports tenants to interactively customize and configure software systems. 
\item RTDDM: The method of supports offline table to drive a data management system development. This method is also coding-free, and enables users to develop and deploy the data management system on the SaaS platform efficiently by directly customizing the system requirements into a given specific table defined by us. Based on this method, we design a scalable and RTDDM-enabled system Requirements Table (ReTa), which can completely describe a software system with data management as the core requirement.
\item GDWI: It is a set of service interfaces in the general Web data management system. 
\end{itemize}

The rest of this paper is organized as follows: Section 2 introduces the system architecture of the proposed framework and the common characteristics model of the data management system. Section 3 introduces the architecture and implementation of DMSWP in detail. Section 4 introduces RTDDM and requirements table ReTa. Section 5 verifies the feasibility and availability of the framework through experiments. Section 6 discusses the related work.  Finally, we summarize the work, and prospect the future research direction.
\section{Framework Achitecture and System Modeling}
This section introduces in detail the modeling process of data management systems with common characteristics, as well as the system architecture and working modules of the framework proposed in this paper.
\subsection{Framework Architecture}
\begin{figure}
\centering
\includegraphics [width=10cm] {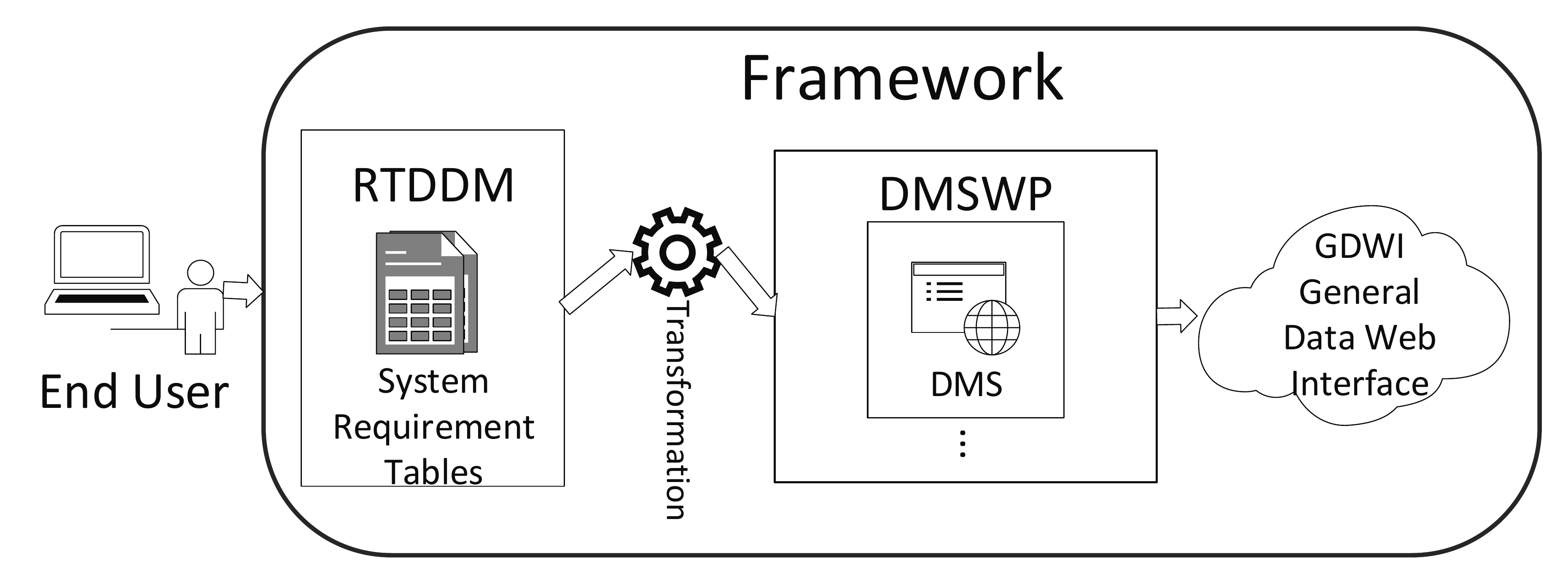}
\caption{The Conceptual Architecture of the Framework}
\label{Fig1ConceptualArchitecture}
\end{figure}

The system architecture is shown in Figure~\ref{Fig1ConceptualArchitecture}. This framework is mainly composed of three parts, RTDDM, DMSWP and GDWI.

DMSWP development platform is the core module of this framework. The platform extracts the core commonness of the data management system and designs a Web framework.

RTDDM is a method of offline customization system using spreadsheet.  In RTDDM module, we design a set of requirements table, which called ReTa. The RTDDM engine will analyze and execute by the system requirements table and derive the corresponding system instance on the DMSWP.

The GDWI provides users with the interface of data services and an operating environment. This module is based on DMSWP and compatible with the RTDDM.
\subsection{Systems Modeling}
Before implementing the framework, it is necessary to design a unified architecture for the data management system. Therefore, in this section, we extract the core commonness of data management system and then proceed system modeling.

We abstract the commonness of the Data Management System (DMS), which is defined with a five-tuple as Definition~\ref{Definition 1}:
\newtheorem*{Definition 1}{Definition 1}
\begin{Definition 1}\textsl{Data Management System}
\begin{equation}
\Psi=F(Tenant, Group, User, Schema, Data)
\label{Definition 1}
\end{equation}

The Tenant is the system administrator who uses the framework to create the system, and we will use the term tenant instead of administrator below. The Tenant describes tenant information in the DMS. The Group describes user group information in the DMS. The User describes user information in the DMS. The Schema describes data structure in the DMS. The Data describes data record items in the DMS. The F represents the engine transformation function. The details on them are given as follows.
\end{Definition 1}
\newtheorem*{Definition 1.1}{Definition 1.1}
\begin{Definition 1.1}\textsl{Tenant Information Representation}
\begin{equation}
T=\{t_i|i=1,2,...,k\}
\end{equation}

The T represents the tenant of data management systems. 
\end{Definition 1.1}
\newtheorem*{Definition 1.1.1}{Definition 1.1.1}
\begin{Definition 1.1.1}\textsl{Tenant Structure Representation}
\begin{equation}
t=(id, password, name)
\end{equation}

The t represents the basic information of a tenant in each system, where id is the unique identifier of the tenant, password is the security authentication password of the tenant, and name is the title of the system.
\end{Definition 1.1.1}
\newtheorem*{Definition 1.2}{Definition 1.2}
\begin{Definition 1.2}\textsl{Group Information Representation}
\begin{equation}
G=\{g_i|i=1,2,...,k\}
\end{equation}

The G represents a collection of user groups contained in a data management system. The division of user groups facilitates the access management of the system. The specific attribute representation of each g is shown in Definition 1.2.1.
\end{Definition 1.2}
\newtheorem*{Definition 1.2.1}{Definition 1.2.1}
\begin{Definition 1.2.1}\textsl{Group Structure Representation}
\begin{equation}
g=(groupid)
\end{equation}

The g represents each group of a data management system. Each g takes groupid as its unique identifier.
\end{Definition 1.2.1}
\newtheorem*{Definition 1.3}{Definition 1.3}
\begin{Definition 1.3}\textsl{User Information Representation}
\begin{equation}
U=\{u_i|i=1,2,...,k\}
\end{equation}

The U represents a collection of users contained in a data management system. The specific attribute representation of each u is shown in Definition 1.3.1.
\end{Definition 1.3}
\newtheorem*{Definition 1.3.1}{Definition 1.3.1}
\begin{Definition 1.3.1}\textsl{User Structure Representation}
\begin{equation}
u=(userid, username, userpwd, G_u)
\end{equation}

The u represents each user of a data management system. The userid is the user's unique identifier in the system. Userpwd is the user's security authentication password when logging in. Username is the user's nickname. G$_{u}$ represents the user group to which the user belongs, which can be one or more.
\end{Definition 1.3.1}
\newtheorem*{Definition 1.4}{Definition 1.4}
\begin{Definition 1.4}\textsl{Schema Information Representation}
\begin{equation}
S=\{s_i|i=1,2,...,k\}
\end{equation}

The S represents a collection of data structure schemas contained in a data management system. Its specific attribute representation is shown in the sub-definition.
\end{Definition 1.4}
\newtheorem*{Definition 1.4.1}{Definition 1.4.1}
\begin{Definition 1.4.1}\textsl{Schema Structure Representation}
\begin{equation}
s=(schemaid, group, entry, gpermission, opermission, FI)
\end{equation}

The s represents each schema of a data management system. Schemaid is the unique identifier of each schema in the system. Group represents schemas belonging to a group of users. Entry represents the user with the highest authority on the schema. Gpermission and opermission will be used to divide the operation permissions of the user belonging to the group from those of other users. FI means the field information contained in the s(schema), namely the description of data structure. FI is defined as follows.
\end{Definition 1.4.1}
\newtheorem*{Definition 1.4.2}{Definition 1.4.2}
\begin{Definition 1.4.2}\textsl{FI Information Representation}
\begin{equation}
FI = \{ fi_{i}\vert i =1,2...k\}
\end{equation}

FI represents the field set of schema, within which there can be multiple fi, each fi is defined as described in the sub-definition.
\end{Definition 1.4.2}
\newtheorem*{Definition 1.4.2.1}{Definition 1.4.2.1}
\begin{Definition 1.4.2.1}\textsl{FI Structure Representation}
\begin{equation}
fi=(fname, ftype, OAs)
\end{equation}

Fname represents field name. Ftype represents the field type. OA means other attributes. The OA can be customized flexibly, for example, it can be customized field properties, whether to allow null.
\end{Definition 1.4.2.1}
\newtheorem*{Definition 1.5}{Definition 1.5}
\begin{Definition 1.5}\textsl{Data Item Representation}
\begin{equation}
D=\{d_i|i=1,2...k\}
\end{equation}

D represents the structured data items contained in a data management system. The specific attribute representation of each u is shown in Definition 1.5.1.
\end{Definition 1.5}
\newtheorem*{Definition 1.5.1}{Definition 1.5.1}
\begin{Definition 1.5.1}\textsl{Data Structure Representation}
\begin{equation}
d=(v_1, v_2,...,v_n)
\end{equation}

The d represents each record of a data management system. The v$_{i}$ means the value of the i-th field.
\end{Definition 1.5.1}

According to the basic characteristics of the defined data management system, we construct the system theorems based on the definitions. The three main basic theorems of the system framework are as follows.
\newtheorem*{Theorem 1}{Theorem 1}
\begin{Theorem 1}\textsl{Uniqueness Theorem}
\begin{equation}
\forall t\in T,\exists !\Psi _{t}=F\left ( t,* \right )
\end{equation}

Where T represents the collection of all tenants. Theorem 1 means that for any given tenant, there is and only one system that can be created corresponding to it, that is, the relationship between the tenants and the constructed systems is one-to-one correspondence.
\end{Theorem 1}
\newtheorem*{Theorem 2}{Theorem 2}
\begin{Theorem 2}\textsl{Isolation Theorem}
\begin{equation}
\forall i\neq j,\Psi _{i}\cap \Psi _{j} =  \emptyset
\end{equation}

Theorem 2 means that any two different systems are completely isolated. The empty intersection means that different systems created by different tenants in the platform are completely unrelated to one another, and there is no content conflict between them.
\end{Theorem 2}
\newtheorem*{Theorem 3}{Theorem 3}
\begin{Theorem 3}\textsl{Expansion Theorem}
\begin{equation}
\forall i\in N> 0,\Omega =\bigcup _{i=1}^{N}\left \{ \Psi _{i} \right \}
\end{equation}

Theorem 3 shows that the number of tenants on the platform can be scalable up to any N.
\end{Theorem 3}
In addition to the three theorems, there are some Constraint Rules (CRs) among the defined system properties. These constraints can be broadly classified into three categories CRs between User and Group, CRs between Schema and Group and CRs between Schema and Data.

{\bf 1)} For CRs between User and Group, a user U belonging to group G has authorities of group G.
\begin{itemize}
\item For any system, there can be 0 or more groups G.
\item For each group G, there can be 0 or more users U.
\item Each user U can belong to one or more groups G.
\end{itemize}

{\bf 2)} For CRs between Schema and Group, a group G has read and write authorities to schema S and schema-derived data D.
\begin{itemize}
\item For any group G, which has the authorities of schema S, there can have authorities of any subset of schema S.
\item Each schema S can be read or written by 0 or more groups G.
\item Any user U's authorities are the union set of the authority of the group G which it belongs to.
\end{itemize}

{\bf 3)} For CRs between Schema and Data, each piece of data D belongs to a schema S.
\begin{itemize}
\item For any system, there can be 0 or more data structure schemas S.
\item For each schema S, there can be 1 or more fields FI.
\item For each field FI, there can be 0 or more field attributes OA.
\item Each piece of structured data D belongs to a schema S.
\item For each structured data D belongs to schema S, the values' attributes of each field must match the attributes of the corresponding schema S's field.
\end{itemize}

\begin{figure} [H]
\includegraphics[height=10cm]{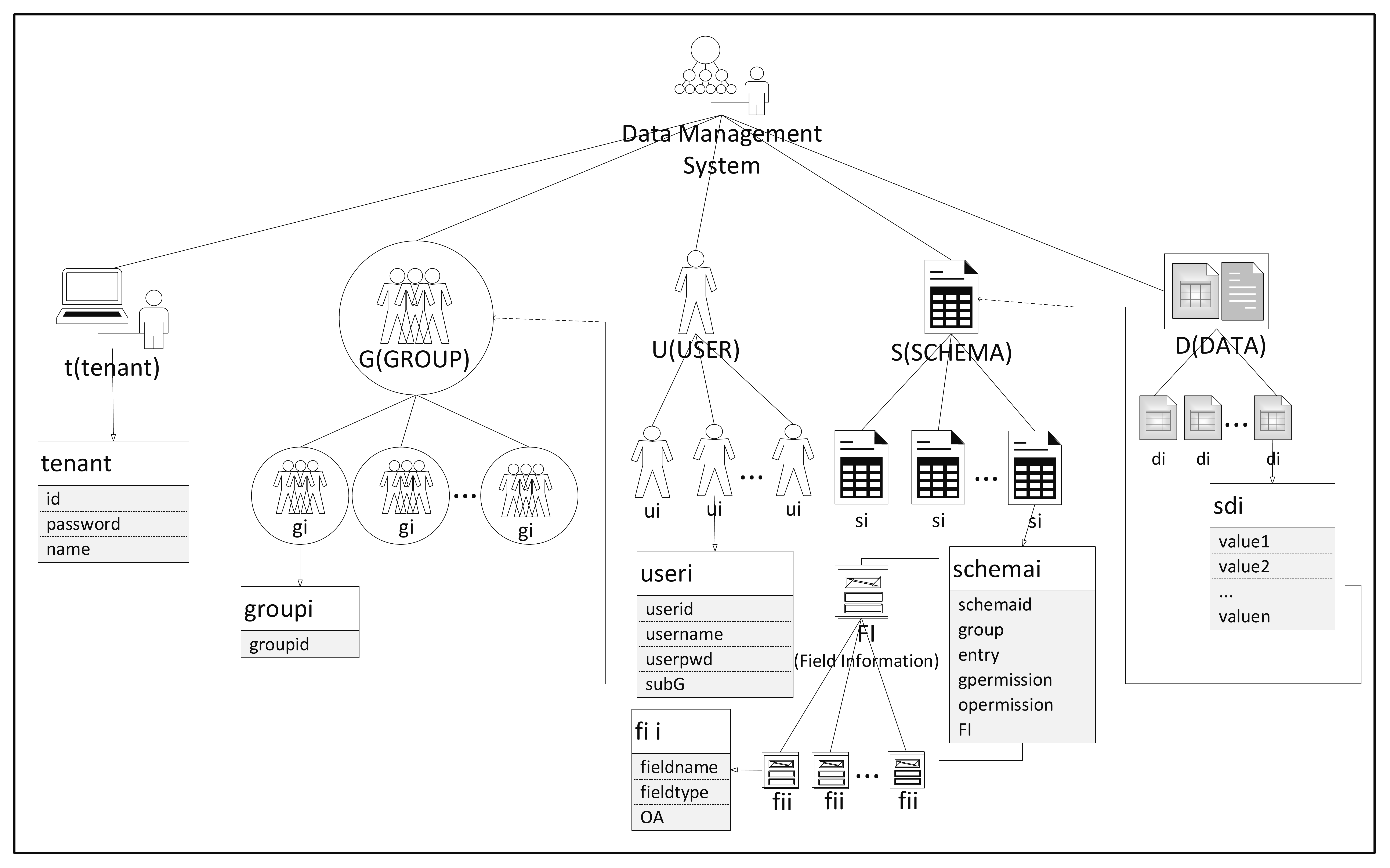}
\caption{The Hierarchy of Data Management System Architecture}
\label{Fig2}
\end{figure}

According to the above-mentioned, the hierarchy of the general data management system's architecture can be represented by Figure~\ref{Fig2}.
\section{Data Management Systems Web Development Platform in the Cloud}
\begin{figure} [H]
\centering
\includegraphics [height=8cm] {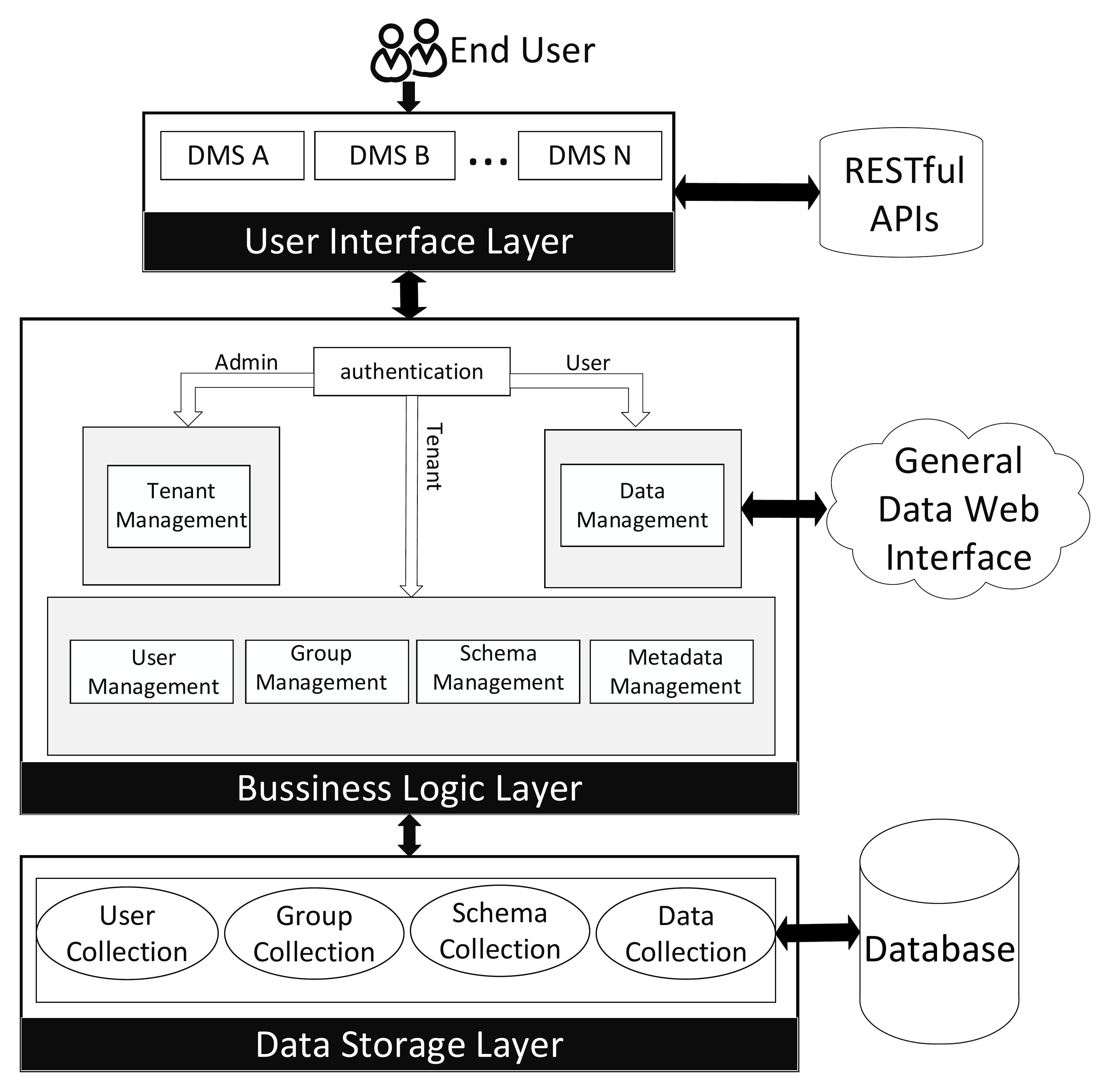}
\caption{DMSWP Deployment Architecture}
\label{Fig3DMSWP_Architecture}
\end{figure}

This section introduces the development platform of data management system based on SaaS, which called DMSWP, that we designed and implemented. Figure~\ref{Fig3DMSWP_Architecture} shows the deployment architecture of DMSWP.

DMSWP is roughly divided into three layers, namely User Interface Layer, Business Logic Layer and Data Storage Layer. The data management of the business logic layer contains the general data web interface GDWI, which can enable accessing the data.

\begin{figure} [H]
\centering
\includegraphics[height=7cm]{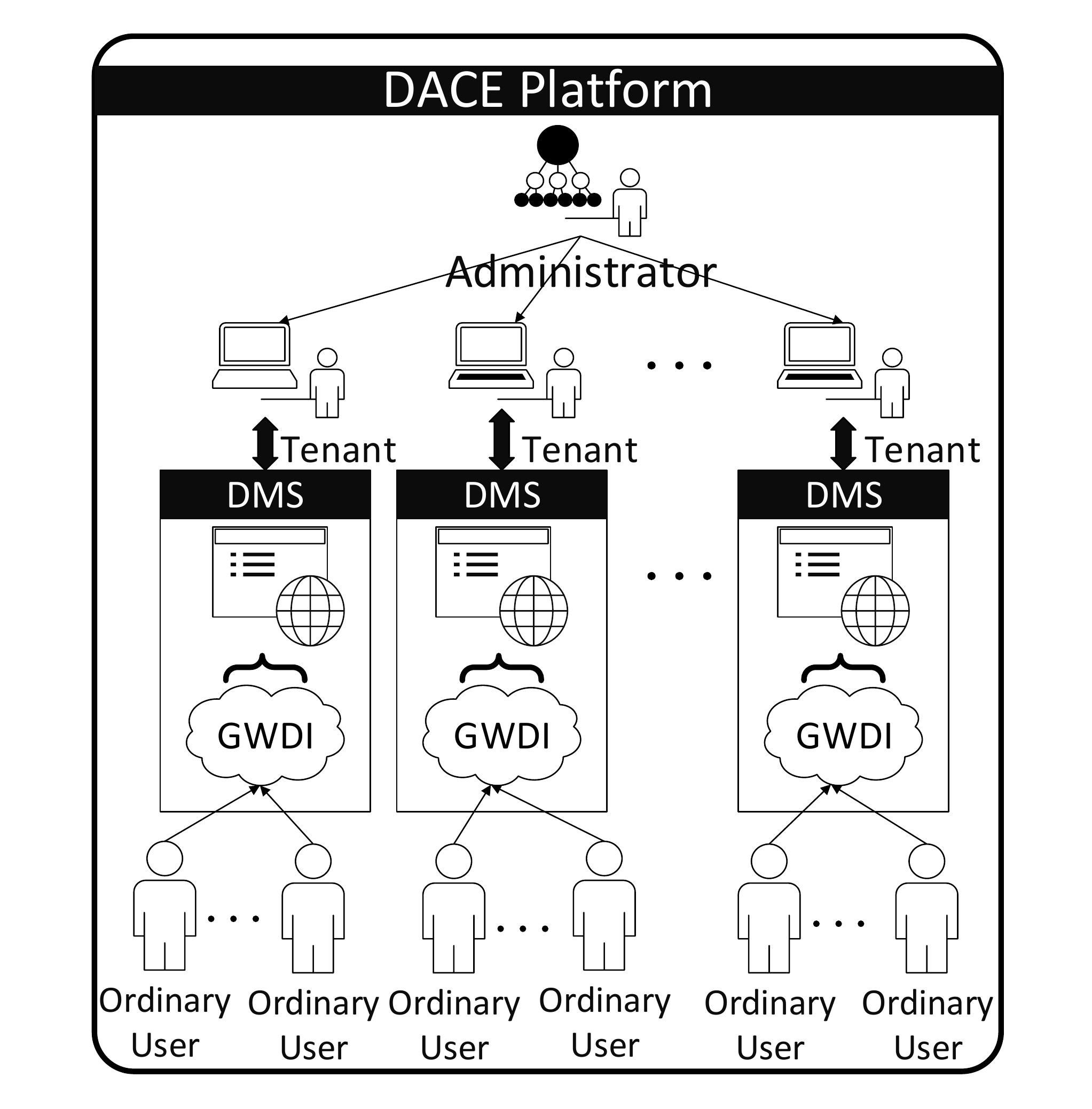}
\caption{DMSWP Role Relationship}
\label{Fig4Roles}
\end{figure}

For DMSWP, end users can be divided into three categories, administrators, tenants and ordinary users. For different categories of users, DMSWP provides different business logic to satisfy the service on demand. For the administrator, the corresponding business logic layer service involves tenant management. That is, the administrator manages the system registered on DMSWP uniformly. For tenants, the corresponding business logic layer services involves user management, group management, schema management and metadata management. For ordinary users, namely users under tenant, the corresponding service of business logic layer is data management. The relationship diagram is shown in Figure~\ref{Fig4Roles}.

To facilitate maintenance and migration, the DMSWP adopts a loosely-coupled design between layers. Next, we elaborate on the three-layer structure.

{\bf User Interface Layer (UIL)} provides customers with access to the data management system on the Internet, and supports the interactive operation between customers and the system. The user interface layer is implemented by Google GWT and SmartGWT, and the separation from the business logic layer is realized through RPC communication. 

{\bf Business Logic Layer(BLL)} mainly includes login verification, tenant management, user management, group management, schema management and data management. For data management module, we also implement GDWI module. In the GDWI module, we implements advanced functions such as multi-condition combined query, statistics and data migration. The data migration function includes import and export. The import function includes import from spreadsheet and traditional relational database such as MySQL. Export function can export the data from data management system into spreadsheet.

\begin{figure}
\centering
\includegraphics[height=12cm]{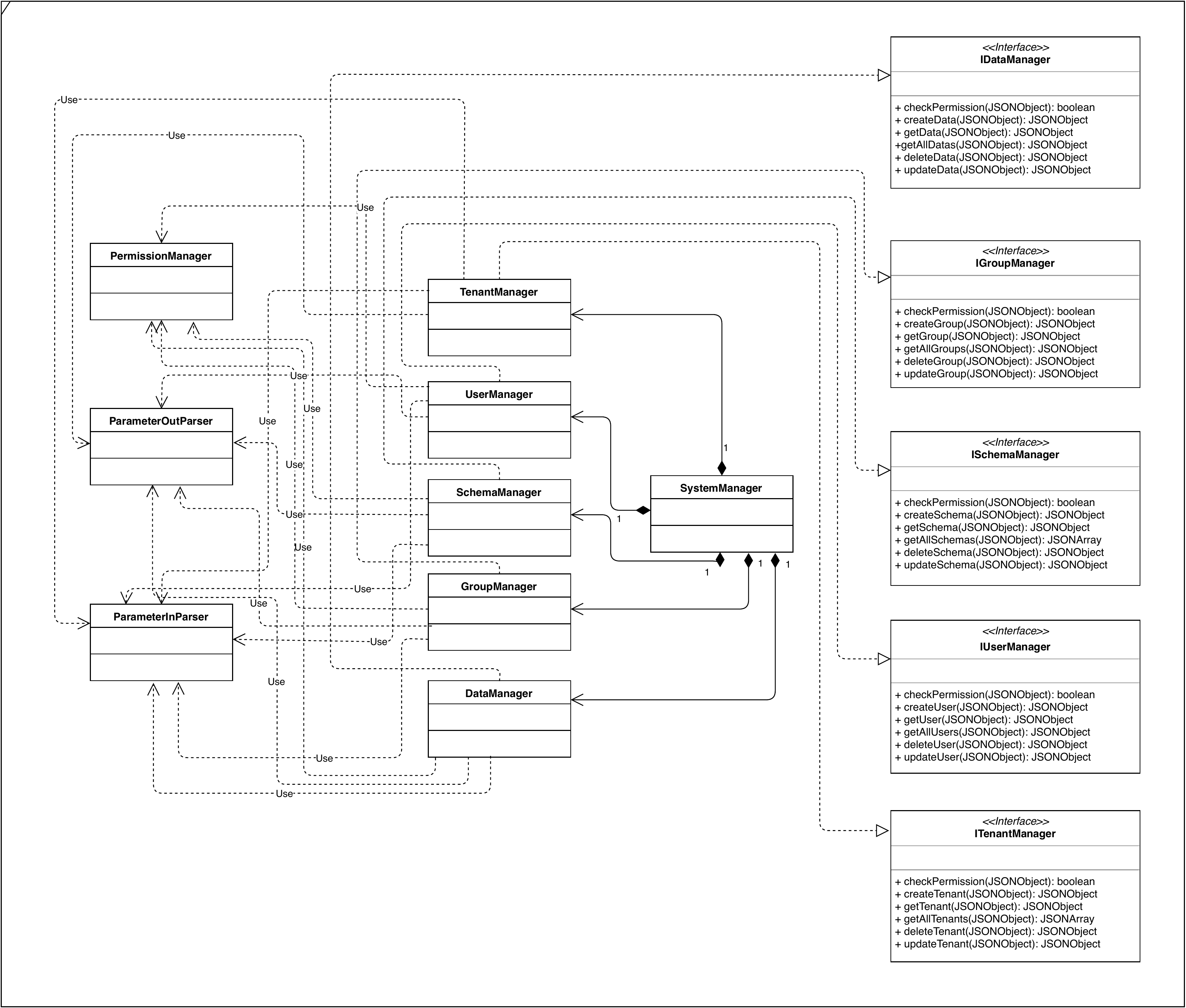}
\caption{Business Logic Layer Conceptual UML Class Diagram}
\label{Fig6BLLCD}
\end{figure}

The BLL implements the respective Manager classes, and defines the I*Manager interface to implement specific operational functions. The parameters passed into the parsing API are accomplished by the ParameterInParser class. The results returned by the parsing API are accomplished by the ParameterOutParser class. The permission verification is accomplished by calling the PermissionManager class. The conceptual UML class diagram for the business logic layer is shown in Figure~\ref{Fig6BLLCD}.

\begin{figure} [H]
\centering
\includegraphics[height=11cm]{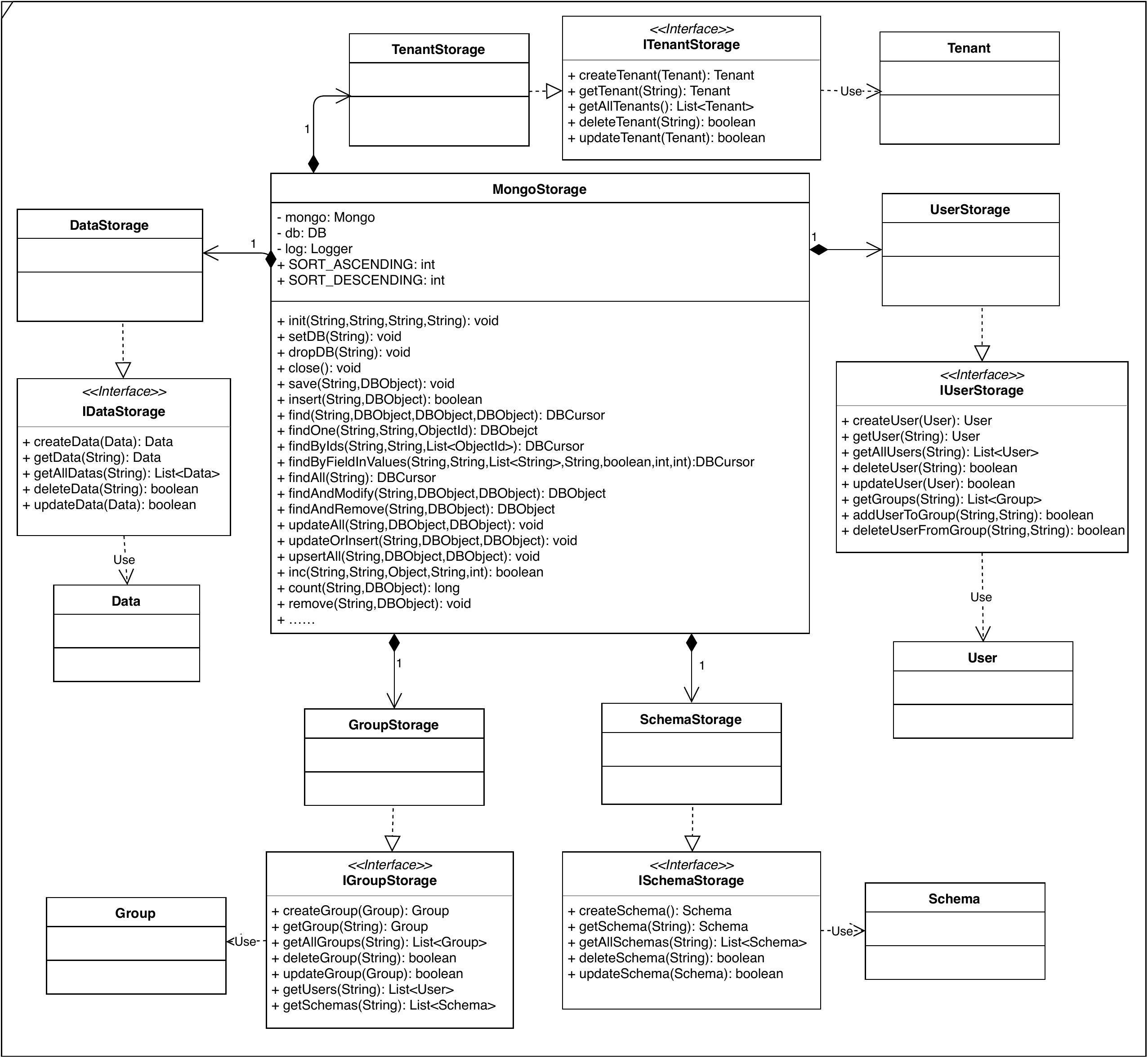}
\caption{Data Access Module UML Class Diagram.}
\label{Fig7DAMCD}
\end{figure}

{\bf Data Storage Layer (DSL)} includes database implementation and unified data access module. In terms of database implementation, the platform adopts the schemaless database MongoDB to establish the database model of multi-tenant application framework. Thanks to MongoDB's schemaless, data of different structured data tables can coexist with the same table. The UML class diagram of data access module is shown in Figure~\ref{Fig7DAMCD}.
\section{Requirements-Table-Driven Development Method}
Based on the characteristics of DMSWP, we propose a development method, called system requirements-table-driven, to develop data management system, and design a set of scalable system requirements table (ReTa).

In this section, we elaborate how to use the system requirements table to describe a system instance on the DMSWP, and describe the workflow for executing this method on DMSWP.
\subsection{On RTDDM Hierarchy}
\begin{figure} [H]
\centering
\includegraphics[height=4cm]{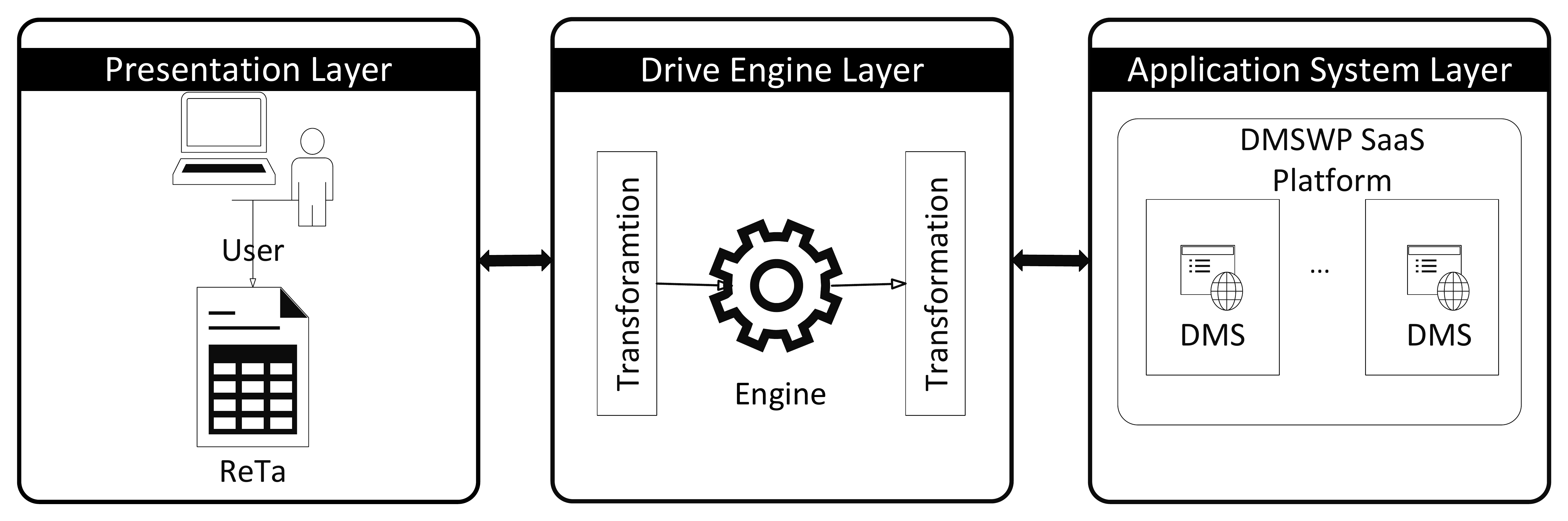}
\caption{RTDDM Workflow.}
\label{Fig8RTDDMworkflow}
\end{figure}

\begin{figure}
\centering
\includegraphics[height=13cm]{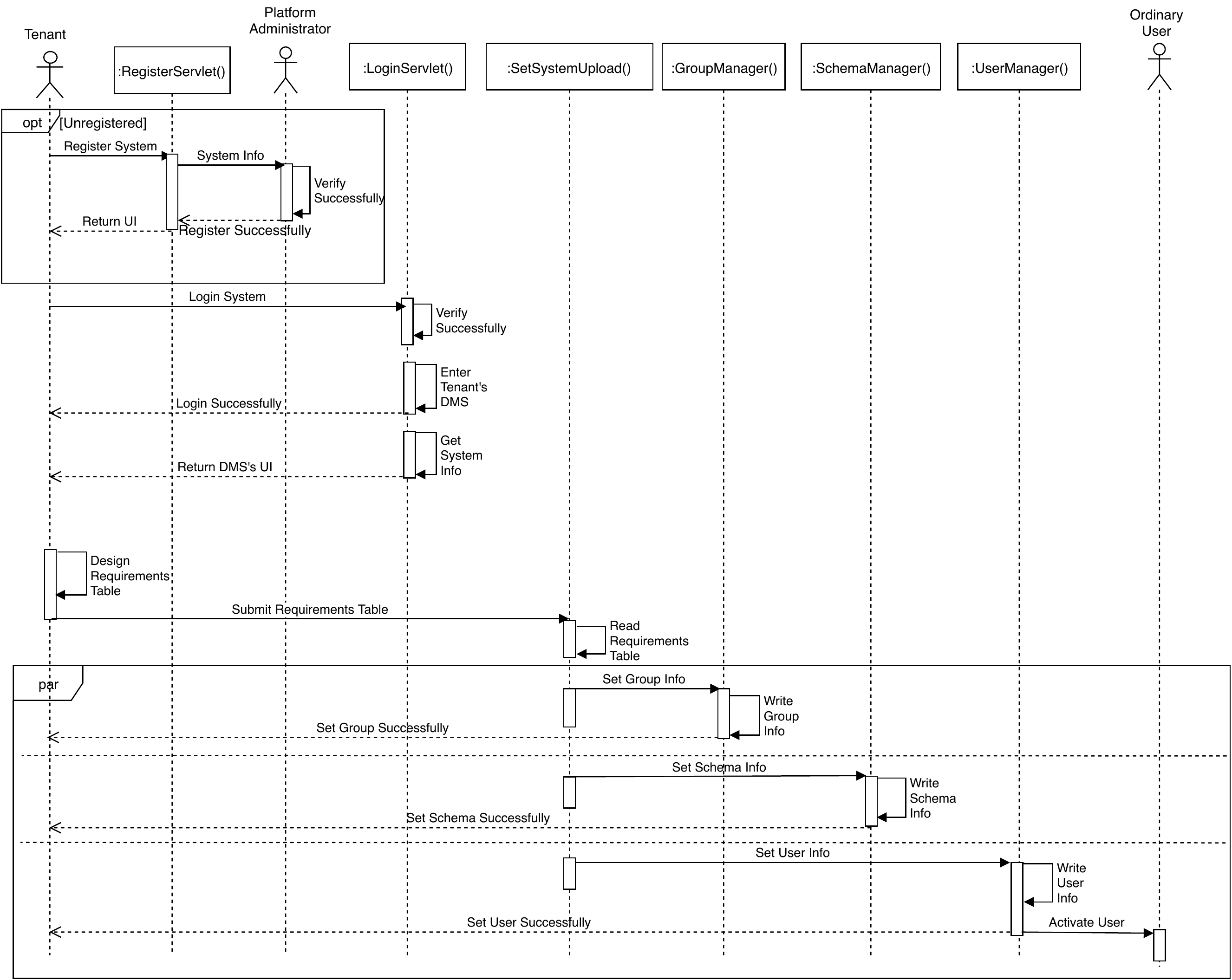}
\caption{RTDDM UML Sequence Diagram.}
\label{Fig9RTDDMSD}
\end{figure}

The RTDDM designed in this paper can customize the personalized data management system through the spreadsheet in specific format. The workflow of the specific RTDDM are shown in Figure~\ref{Fig8RTDDMworkflow}. As shown in the figure, RTDDM mainly includes three layers, namely Presentation Layer, Drive Engine Layer and Application System Layer. 
Figure~\ref{Fig9RTDDMSD} is the sequence diagram of users using RTDDM. It describes how users can develop a system through RTDDM. After users register and log in, the system can be generated by uploading spreadsheet files. Then users can manage the system by DMSWP.
\subsubsection{Presentation Layer}
The presentation layer is the layer where the user interacts directly with the RTDDM. In RTDDM, the presentation layer takes the well-known spreadsheet as the input form, so that users have a convenient and simple entity to interact with the DMSWP. The presentation layer describes users' requirements of a data management system by designing a set of system requirements tables. Users configure the system requirements table to customize the data management system. The presentation layer in the RTDDM assembles services on demand from a user perspective. A structure of the system requirements table is shown in section~\ref{subsec:RT}.
\subsubsection{Drive Engine Layer}
\begin{figure}
\centering
\includegraphics[width=10cm]{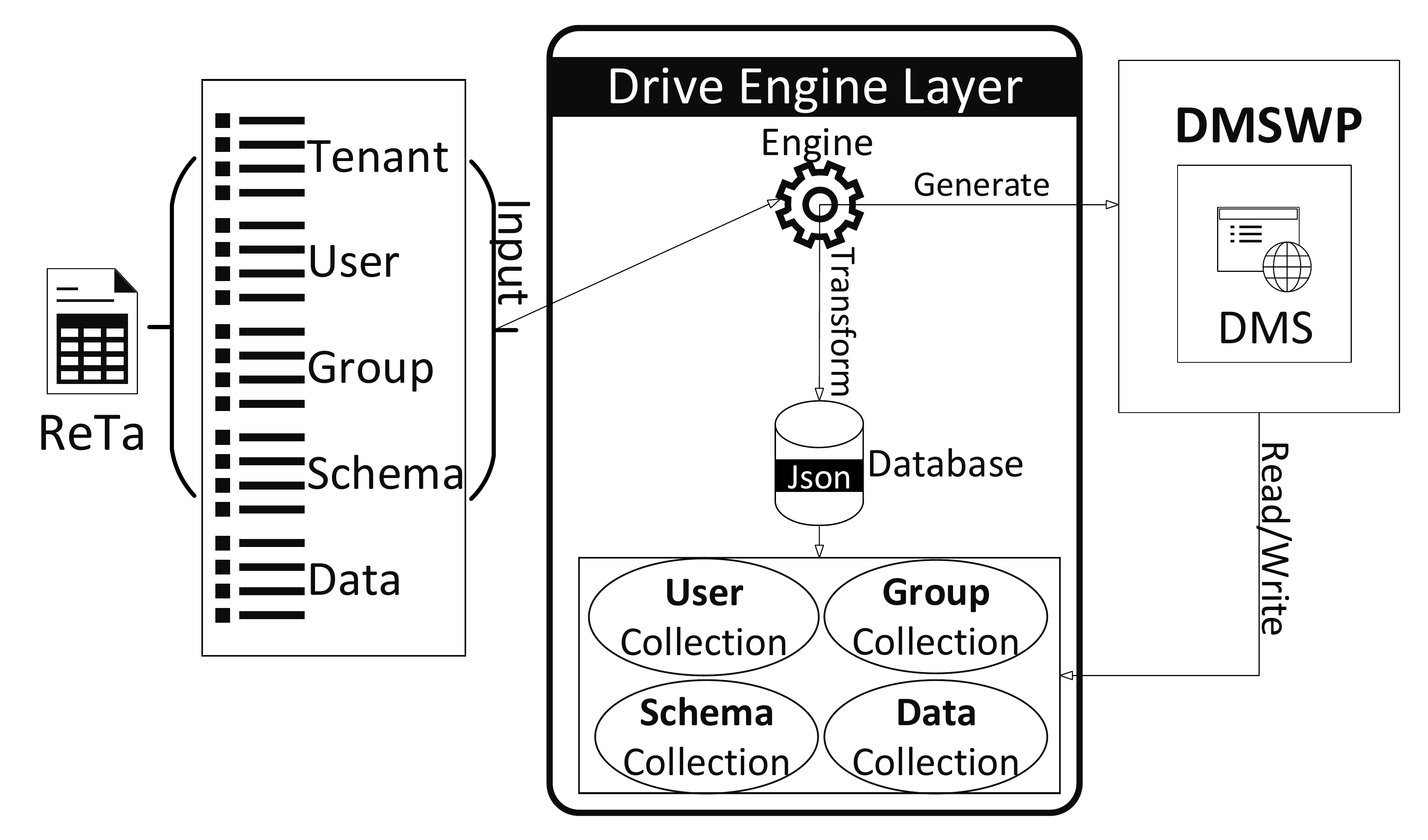}
\caption{Drive Engine Layer Process.}
\label{Fig10DELP}
\end{figure}

The drive engine layer is the bridge that converts the system requirements table from the presentation layer to the system instances on the application system layer. The main task of the drive engine layer is to parse the system requirements table customized by the tenant. The system requirements table passed from the presentation layer is read, and the information is verified. After passing the verification, it is interpreted into the language that can be parsed and executed on the DMSWP. The specific work process of the drive engine layer is shown in Figure~\ref{Fig10DELP}.
\subsubsection{Application System Layer}
The application system layer is in charge of the instantiation of the system. The data management system customized by the user in the system requirements table is transformed and parsed by the drive engine layer, executed on the DMSWP, and then generated and derived into the web. Users can access to the customized system on the web.
\subsection{On Requirement Tables}
\label{subsec:RT}
The RTDDM proposed in this paper extracts the common characteristics of the data management system, and then models the system. We materialize the data requirements table model, propose the tables to construct a system, and describe the model of the data management system. RTDDM's table includes two types of data requirements table. One is a system metadata table, and the other is a data exchange table.
\subsubsection{System Metadata Table}
The system metadata table has a detailed description on users, groups and schemas in the data management system. The format of the system metadata table is shown as Table ~\ref{Tab3SysMetaT}.
\begin{table} [H]
\centering
\caption{System Metadata Table.}
{\begin{tabular}{|l|c|c|c|c|c|c|c|c|}
\hline
1&T&id&password&systemName\\
\hline		
2&+\\
\hline					
3&G&groupid\\
\hline				
4&+\\
\hline					
5&U&userid&username&userpwd&subG\\
\hline	
6&+\\
\hline					
7&S&schemaid&groupid&entry&gpermission&opermission\\
\hline
8&+\\	
\hline				
9&FI&ftype&fname&F$_{oa}$\\
\hline	
10&+\\				
\hline
\end{tabular}}
\label{Tab3SysMetaT}
\end{table}
In this table, Row1 and Row2 are used for security verification of tenant T. Row3 and Row4 describe the group G of the data management system, and incrementing the cell to the right to add the group. Row5 begins describing user U of the system, one user record per line. Row7 and Row8 describe the basic information of schema S of the system. Row9 begins to describe the field information FI within the schema. User can continue to extend down the table to add a new model or field. The symbol ``+'' in the table means supporting extension. In fact, for the entire platform, each content following with ``+'' can be extended indefinitely as long as the physical space is enough.
\subsubsection{Data Exchange Table}
Data exchange table gives out a detailed description of structured data D in a data management system. The format of the data interchange table is shown in Table~\ref{Tab3DataExchangeT}.
\begin{table} [H]
\centering
\caption{Data Exchange Table.}
{
\setlength{\tabcolsep}{5mm}{
\begin{tabular}{|l|c|c|c|c|c|c|c|c|}
\hline
&fi$_{1}$  &fi$_{2}$  &...&fi$_{k}$    \\
\hline		
1&v$_{11}$&v$_{12}$&...&v$_{1k}$\\
\hline					
2&v$_{21}$&v$_{22}$&...&v$_{2k}$\\
\hline				
...&...&...&...&...\\
\hline					
n&v$_{n1}$&v$_{n2}$&...&v$_{nk}$\\
\hline
\end{tabular}}}
\label{Tab3DataExchangeT}
\end{table}
In each data exchange table, we can only inject data items within the same data structure. The first row of each data exchange table is a description of the data structure, the other rows represent a piece of data record that conforms to the data structure, and each column corresponds to a field.
\section{Evaluation}
\subsection{Design}
In order to verify the effectiveness in terms of simplifying development process and improving development efficiency of development method proposed in this paper, our investigation and evaluation experiment adopted three methods to develop data management system, and conducted three groups of experiments.
\begin{itemize}
\item \textbf{Mode 1:} Use the conventional software development mode to develop a data management system by programming code.
\item \textbf{Mode 2:} Use DMSWP to develop and deploy a data management system by online interaction mode. 
\item \textbf{Mode 3:} Use RTDDM to develop and deploy a data management system by offline ReTa mode.
\item \textbf{Experiment 1:} Develop and deploy a DMS in Mode 1.
\item \textbf{Experiment 2:} Develop and deploy a SaaS DMS in Mode 2.
\item \textbf{Experiment 3:} Develop and deploy a SaaS DMS in Mode 3.
\end{itemize}
\subsection{Conduct}
\subsubsection{Participators}
We recruited 20 users with programming knowledge background and assigned them to group A (programmer group). Most of them are good at computer programming. The average programming age of group A is about 2.5 years. At the same time, we also recruited 20 users without programming experience and assigned them into group B (non-programmer group).
\subsubsection{Teaching}
We gave simple guidance on the use of DMSWP and RTDDM for both groups A and B. We have prepared a case of using RTDDM to quickly generate a software system on DMSWP.  Fill in the requirements according to the format of Table ~\ref{Tab3SysMetaT}, and then upload the table to DMSWP. Upon successful upload, the corresponding software system can be generated immediately. In addition, ordinary users created by the system can log in the system to manage their authorized schema data. 
\subsubsection{Tasks Requirements}
We assigned the 40 users the task of developing and deploying a common vehicle management system in three different ways. The system requirements are as follows:
\begin{itemize}
\item There are at least two types of users, namely customers and supervisors.
\item There are at least three types of information, namely vehicle information, maintenance information and customer information.
\item A different type of users has different authorities to different information. Tenants must have the authority to ADD, DELETE and UPDATE the system's metadata. Ordinary users must have the authority to operate the data, and different ordinary users with different authorities must have different operation permissions.
\item The development process prohibits the assistance of others who are familiar with programming knowledge.
\end{itemize}
\subsection{Result}
\begin{figure}
\centering
\subfigure[Completion Rate]{
\begin{minipage}[b]{0.45\linewidth}
\includegraphics[width=7.2cm]{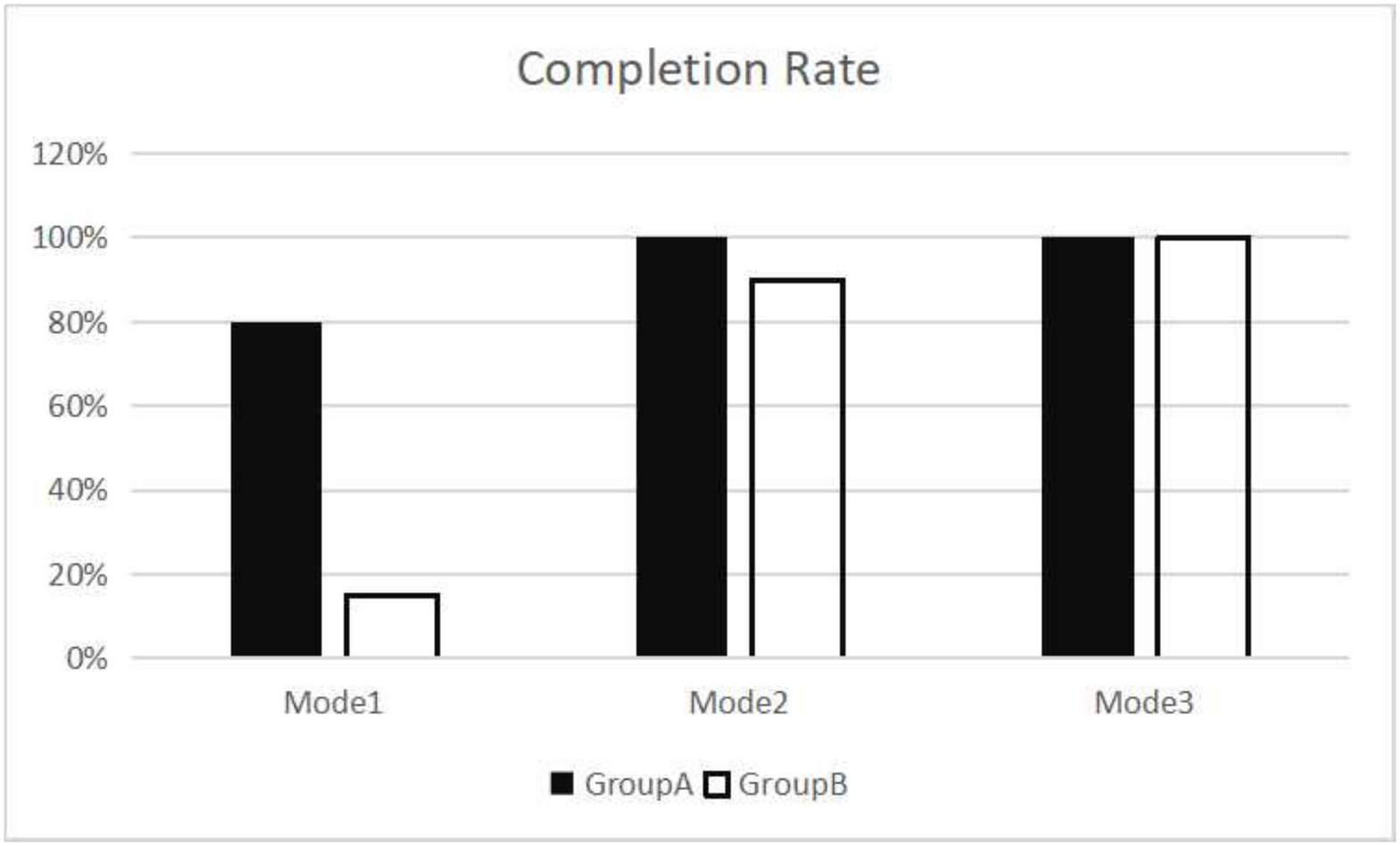}
\end{minipage}}
\subfigure[Development Time Cost]{
\begin{minipage}[b]{0.45\linewidth}
\includegraphics[width=7.2cm]{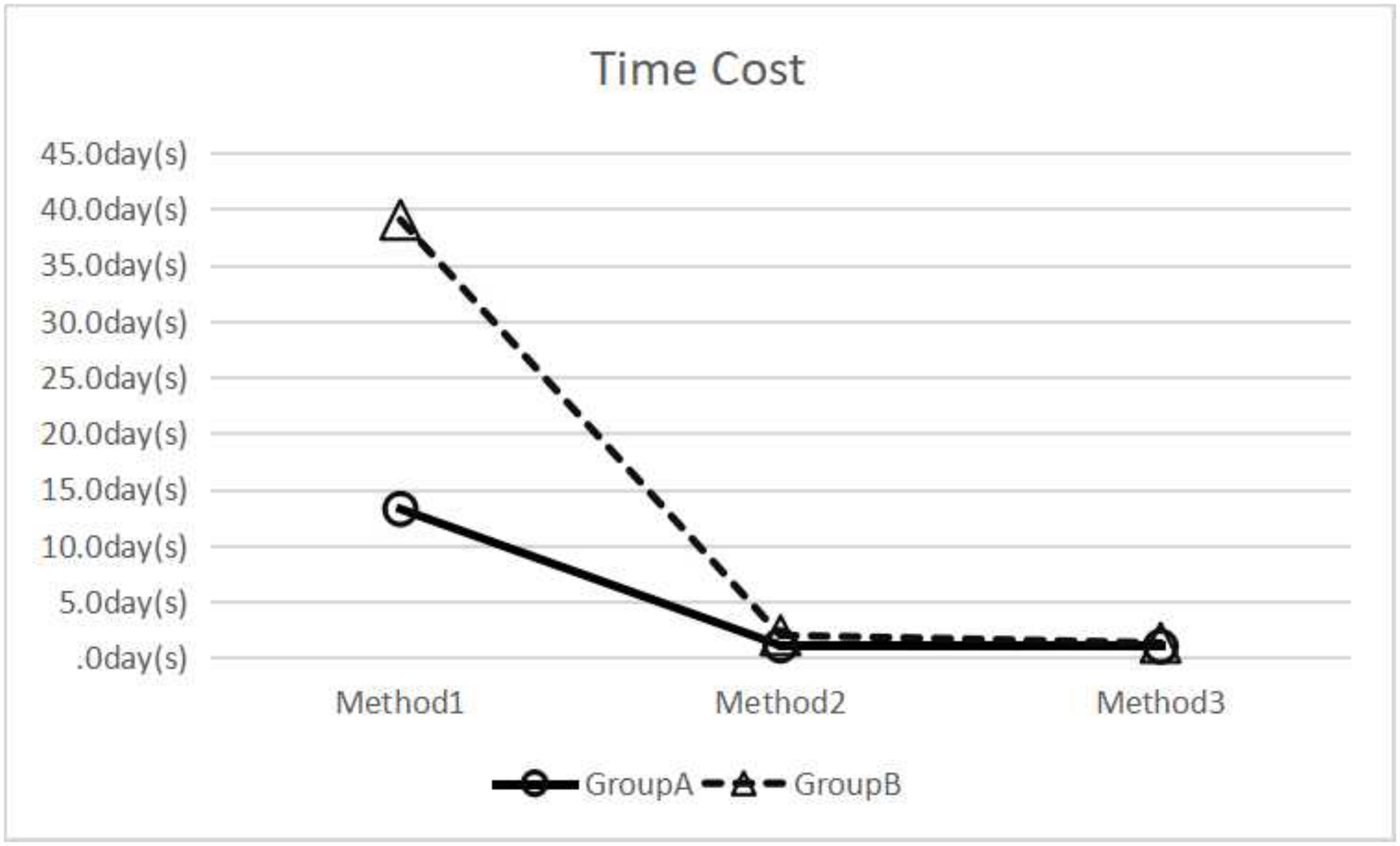}
\end{minipage}}
\caption{Different Groups with Different Modes Experiment Result.}
\label{Fig11result}
\end{figure}

The comparative analysis of experimental results is shown in Figure~\ref{Fig11result}.
\subsubsection{Results of Group A}
\begin{itemize}
\item Using \textbf{Mode 1}, 16 people completed the task, with an average development time of 13.3 days. The fastest users completed in 2 days, while the slowest users completed in 15 days.
\item Using \textbf{Mode 2}, all members completed the development. Only 2 users spent more than 1 day, while the remaining 18 users spent less than 1 day, with an average development time of 1.1 days. The development time of all the 16 users who completed the task with Mode 1 was less than 1 day.
\item Using \textbf{Mode 3}, all members completed development, and the development time of all 20 users was less than 1 day.
\end{itemize}

We compared the completion rate and completion time respectively. The development task completion rate of group A in three ways is shown in Figure 11(a), which is shown in the solid column. The development time of the 16 users who completed the development in all three ways is shown in Figure 11(b), which is shown by the hollow circle line.
\subsubsection{Results of Group B}
\begin{itemize}
\item Using \textbf{Mode 1}, only 3 users finally completed the task. From the day of basic teaching, the average development time of the 3 users was 39 days.
\item Using \textbf{Mode 2}, 18 people completed the development, with an average development time of 2.94 days. The average development time of the 3 users who completed the development with Mode 1 was 2 days when Mode 2 was used.
\item Using \textbf{Mode 3}, all users completed the development, with an average development time of 1.85 days. 18 users who completed development in Mode 2 had an average development time of 1.78 days when using Mode 3. The average development time for the 3 users who completed the development in Mode 1 was 1.33 days when Mode 3 was used.
\end{itemize}

This was a daunting task for the non-professional users in group B. We compared the completion rate and completion time respectively. The development task completion rate of group B in three ways is shown in the hollow column in Figure 11(a). The development time of three users who have completed the development in all three ways is shown in Figure 11(b), which is shown in the hollow triangle dotted line.

The above experimental data show that for group A, the development time of \textbf{Mode 2} is 12.2 days less than that of \textbf{Mode 1}. Using \textbf{Mode 3}, the development time is reduced 12.3 days compared to \textbf{Mode 1}. For group B, the development time is reduced 37 days via using \textbf{Mode 2} compared to \textbf{Mode 1}. Using \textbf{Mode 3}, the development time is reduced 38 days compared to \textbf{Mode 1}. Experiments show that the DMSWP and RTDDM have a great achievement in simplifying the development process, improving development efficiency and completion rate.
\section{Related Work}
We survey the literature on SaaS and spreadsheets, related to software development.

\subsection{System Software Development based on SaaS}
In recent years, a growing number of professionals have discussed how to develop and deploy SaaS applications efficiently and have focused on modeling SaaS applications.

At present, several researchers have proposed different technologies for customizable SaaS platforms. Tsai, Huang, and Shao\cite{Tsai11} proposed a SaaS development framework designed to simplify SaaS development, EasySaaS. Moreover, Tsai et al.\cite{Tsai12} proposed a model-driven SaaS application platform to develop a complete application through the model description file, providing a convenient and quick solution for program customization. Further more, Tsai, and Sun\cite{Tsai13} proposed an innovative customization approach that studies similar tenants' customization choices and provides guided semi-automated customization process for the future tenants. They divided SaaS applications into four layers GUI, business logic, service and data. Different aspects of the application are customized and modeled using OVM, and a guided framework was proposed. However, they did not focus on the scalability of the platform. Zhang et al.\cite{Zhang18} implemented a big data intelligent SaaS platform, which supports the dynamic construction and operation of intelligent data analysis applications. In addtion, they used the technology of cloud workflow to realize the rapid development and process deployment of business analysis process.

Besides, most of researches on SaaS software modeling development adopt the Model Driven Architecture (MDA). MDA is a process in which the model is the core and the development is driven by the model mapping. Mohamed et al.\cite{Mohamed17} proposed an integration platform based on service orientation, Software Product Lines(SPLs) and MDA. This platform supports configuration customization for runtime tenants. Sharma, and Sood\cite{Sharma11} explored the MDA approach for developing SaaS applications. They proposed a development method from PIM (Platform Independent Model) to PSM (Platform Specific Model), then binding from PSM to execution platform. Cai et al.\cite{Cai18} combined MDA ideas to promote SaaS application development. They proposed a model-driven development model based on semantic reasoning mechanisms. However, their work is mostly focused on Cloud of Things(CoT) applications. Ma, Yang, and Abraham\cite{Ma12} combined with the Model Driven Engineering (MDE) idea, proposed a data middleware to customize multi-tenant database, generated SaaS application from the model with the help of model transformation, and rapidly generated application code through model modelings.  However, MDA still needs to master professional techniques and grammars, and it is still difficult for non-professionals to quickly produce software systems through MDA.

In addition, serveral researchers have proposed different ways to quickly customize software on SaaS platforms. Mietzner, Leymann, and Unger\cite{Mietzner10} proposed a multi-tenant pattern that can be used to design, develop, and deploy service-oriented applications that support processes. They described how different SaaS application components can be combined horizontally or vertically. Scheibler, Mietzner, and Leymann\cite{Scheibler09} proposed a platform-independent approach to modeling, describing and implementing EAI patterns in a service-oriented architecture to integrate different applications into a SaaS environment.  Liu et al.\cite{Liu10} and Zhang, Zhang, and Liu\cite{Zhang13} used WSCL to build SaaS services and describe SaaS models. Zhu, and Wang\cite{Zhu09} combined MDA thoughts with SOA, proposed a STML based method for describing business requirements, and implemented integrated development tools to enable users to quickly build SaaS applications. Zhang et al.\cite{Zhang12} proposed a requires-based service-oriented model-driven architecture approach, which applies MDA meta-level analysis to facilitate the construction of SaaS mashup applications. Etedali et al.\cite{Etedali17} proposed a method to support multi-tenant SaaS system with features modeling and XML filtering technology. Their main work was to realize the automatic encoding of XML documents into the format required by Yfilter technology and improve the efficiency of developing and deploying SaaS applications. However, many of their researches are unable to avoid tools that use the C/S architecture in modeling.

Xu et al.\cite{Xu17} proposed a demand-driven data-centric Web service configuration method, which represented data resources as data-centric Web services. Moreover, they implemented a scalable dynamic platform, which enabled people to automatically package information resources as Web services and expose them at the business level. The information resources of different tenants are isolated through REST services. They also modeled the system first and then mapped the model onto the platform. Unlike the DMSWP, our proposed framework focuses more on the logic at the data operation level, while they focused on the logic at the business level. In addition, the requirements-driven method we proposed is easier to understand and use by users without programming background. To the best of our knowledge, no similar method to RTDDM is presented in prior literature.

Jiang et al.\cite{Jiang13} proposed a new multi-tenant architecture design pattern, model-driven layered architecture (LMDA), which divides the model into four layers, namely domain independent component layer, domain related component layer,  abstract business layer and abstract SaaS application layer. They defined detailed language and compliance rules for each layer and define access dependencies between layers and bottom-up components. They also suggested a way to build SaaS tenant applications using each layer model. Comparing LMDA\cite{Jiang13} and the above MDA-driven SaaS system construction method with our work, the RTDDM proposed in this paper supports online and rapid installation and deployment of on-demand services and heterogeneous assembly. However, the assembly environment in LMDA and most of researches above requires specific plug-ins, while our proposed approach requires only commonly used spreadsheets, which are faster, more convenient and more secure than plug-ins.
\subsection{Spreadsheet-driven Development}
Spreadsheet is an information management tool widely used by end users \cite{Benson14}. Researches show that spreadsheet is an information management tool easily accessible and familiar to end users \cite{Hoang11}. Many researchers have come up with ideas in the field of research using spreadsheets to drive development. Benson, Zhang, and Karger\cite{Benson14} provided a system for creating basic Web applications using spreadsheets instead of servers, and described the client-side UI using HTML. The result was a responsive read-write compute Web page. Chang, and Myers\cite{Chang14} proposed the Gneiss system, which tightly integrated the spreadsheet editor with the Web interface builder, so that users could demonstrate the binding between Web GUI elements and cell attributes in spreadsheets when dealing with actual Web service data. Kongdenfha et al.\cite{Kongdenfha09} proposed a spreadsheet based Web mashup development framework. Moreover, Hoang et al.\cite{Hoang10,Hoang11} proposed a universal MashSheet based on spreadsheet, which enabled users to build mashup using spreadsheet examples. Honkisz, Kluza, and Wisniewski\cite{Honkisz18} proposed a presentation approach using spreadsheets as middleware for transforming natural language text into business process models. In similar vein, Wisniewski et al.\cite{Wisniewski19} proposed a way to describe business processes through spreadsheets. They model BPMN processes through spreadsheets, mapping process model elements to a spreadsheet representation. However, they mainly focused the modeling of business processes, while our work focuses on the modeling of data management systems with specific application scenarios. In addition, our platform has been implemented and applied in the real world, but most of previous work has still stayed in researches.

From the existing studies, the main problem of the spreadsheet driven method is how to interactively connect the spreadsheet with the Web. According to the researches, there is no literature using the requirement table-driven development method proposed in this paper to build the data management system so far.
\section{Conclusion and Future Work}
In order to meet today's enterprise requirements to migrate data management system from the local to the cloud, we provide a framework to develop and employ the data management system online or offline. This paper proposes a Web data management system development platform based on SaaS. Moreover, this paper provides an online development method on platform and an offline development method by using requirements tables in spreadsheet. The two methods provided in this paper have no need of programming, which makes the software development be easier and agiler than before. At present, all the platform, method and interface proposed in this paper have been realized and deployed on the could.

In the future, we will further develop custom configurations focusing on business logic based on the framework. At present, we have only implemented the offline drive in the form of a spreadsheet. We are still exploring more convenient and friendly ways to generate system. In addition, we will further implement the ``machine learning as a service'' concept and design a machine learning Web interface compatible with the GDWI interface. We will encapsulate machine learning algorithms, optimizers and so on, which aims to simplify the machine learning process, and provide users with channels for advanced data analysis on the framework. 
\bibliographystyle{unsrt}  


\end{document}